\documentclass[10pt,fleqn,aps,prd,preprintnumbers,showpacs,showkeys,nofootinbib,superscriptaddress,floatfix,tightenlines]{revtex4-1}
\usepackage[utf8]{inputenc}
\usepackage{array}
\usepackage{float}
\usepackage{bm}
\usepackage{multirow}
\usepackage{amsmath}
\usepackage{amsthm}
\usepackage{amssymb}
\usepackage{graphicx}
\usepackage{rotfloat}
\usepackage{amsfonts}
\usepackage{amscd}
\usepackage{amsxtra}
\usepackage{epstopdf}
\usepackage{color}
\usepackage{dcolumn}
\makeatletter
\providecommand{\tabularnewline}{\\}

\usepackage[normalem]{ulem}
\usepackage[dvipsnames]{xcolor}
\usepackage{array}
\usepackage{slashed}
\renewcommand{\sout}{\bgroup \color{red} \ULdepth=-.5ex \ULset}

\makeatother

\begin{document}
\preprint{INHA-NTG-02/2020}
\title{Isospin mass differences of singly heavy baryons}
\author{Ghil-Seok Yang}
\affiliation{Department of Physics, Soongsil University, Seoul 06978, Republic
of Korea}
\email{ghsyang@ssu.ac.kr}

\author{Hyun-Chul Kim}
\affiliation{Department of Physics, Inha University, Incheon 22212, Republic of
Korea}
\email{hchkim@inha.ac.kr}

\affiliation{School of Physics, Korea Institute for Advanced Study (KIAS), Seoul
02455, Republic of Korea}
\begin{abstract}
We study the isospin mass differences of singly heavy baryons, based
on a pion mean-field approach. We consider both the electromagnetic
interactions and the hadronic contributions that arise from the mass
difference of the up and down quarks. The relevant parameters have
been already fixed by the baryon octet. In addition, we introduce the
strong hyperfine interactions between the light quarks inside a chiral
soliton and the Coulomb interactions between the chiral soliton and a
heavy quark. The numerical results are in good agreement with
the experimental data. In particular, the results for the neutral mass
relations, which contain only the electromagnetic contributions, are
in remarkable agreement with the data, which implies that the pion
mean field approach provides a good description of the singly heavy
baryons. 
\end{abstract}
\maketitle

\section{Introduction}
Since W. Heisenberg proposed isospin symmetry to deal with the proton
and the neutron on an equal footing~\cite{Heisenberg:1932dw}, isospin
symmetry has played an essential role in classifying the identities
of newly found
hadrons~\cite{GellMann:1953zza,Gell-Mann:1956iqa,Nishijima:1955gxk}.  
The breaking of isospin symmetry is attributed to two difference sources:
one comes from the electromagnetic (EM) interaction
~\cite{Feynman:1954xxl}, which however caused several conflicts with
experiments such as the mass difference between the proton and the 
neutron~\cite{Coleman:1961jn,Coleman:1964zz,Harari:1966mu,Zee:1971df}. 
After the advent of quantum chromodynamics, the current-quark mass
difference of the up and down quarks~\cite{Lane:1976ea,Gasser:1982ap}
was known to the other source of isospin symmetry breaking. To explain
the isospin mass differences of singly heavy baryons, one has to
consider these two contributions in a consistent manner.  
The masses of singly heavy baryons have been extensively investigated
within various different theoretical approaches well over
decades~\cite{Franklin:1975yu,   Itoh:1975rf, Deshpande:1976vn,
  Ono:1977xy, Wright:1978xi,   Chan:1976iv, Lichtenberg:1977mv,
  Isgur:1979ed, Chan:1983pm,   Chan:1985ty, Tiwari:1985ru,
  Tornqvist:1985se, Hwang:1986ee,   Capstick:1987cw, Itoh:1988xs,
  Verma:1988gg, Sinha:1989dfa,   Cutkosky:1993cc, Franklin:1995hc,
  Varga:1998wp,   SilvestreBrac:2003kd,
  Durand:2005tb, Karliner:2008sv, Guo:2008ns,   Hwang:2008dj,
  Karliner:2019lau}. While most of these works were based on variants of the
constituent quark models with phenomenological inter-quark 
interactions, the MIT Bag model~\cite{Hwang:1986ee} and 
chiral perturbation theory~\cite{Guo:2008ns} were also used to compute
the isospin mass differences of singly heavy baryons. As shown in a
seminal paper~\cite{DeRujula:1975qlm}, the masses of hadrons have been
described based on various different \emph{strong} and
\emph{EM} quark-quark interactions. This idea was also
applied to the description of the isospin mass differences of singly
heavy baryons.  The quark-quark interactions in describing theses
differences consist of the chromomagnetic, EM hyperfine, and Coulomb  
interactions. Thus, it is inevitable to introduce several parameters
to be fixed. However, it is of great importance to reduce
uncertainties arising from these parameters such that the isospin mass
differences of singly heavy baryons should be predicted consistently
both in the charmed and beauty sectors.

Recently, a pion mean-field approach has been developed to explain
various properties of singly heavy baryons~\cite{Yang:2016qdz,
  Kim:2017jpx, Kim:2017khv, Kim:2018xlc, Yang:2018uoj, Kim:2018nqf,
  Kim:2019rcx, Yang:2019tst, Kim:2019wbg}. This approach has a great
virtue that the light and heavy baryons are investigated within the
same framework without introducing almost no free parameter. For
example, it was shown in Ref.~\cite{Yang:2016qdz} that all the
parameters can be taken from those already determined in the light
baryon sector. The only parameter that was introduced additionally was
the heavy quark hyperfine interactions to remove the spin degeneracy
in the baryon sextet. The principal idea of the pion mean-field
approach was first proposed by Witten~\cite{Witten:1979kh,
  Witten:1983}. In this approach, a singly heavy baryon can be viewed
as a state consisting of $N_c-1$ valence quarks bound by the pion mean
fields, which are created in the presence of the $N_c-1$ valence
quarks self-consistently,  while a heavy quark inside a singly heavy baryon
is considered to be a mere static color source in the limit of the
infinitely heavy quark mass. The SU(3) representations of singly heavy
baryons, i.e., the baryon antitriplet and sextet naturally appear in
the pion mean-field approach.

In the present work, we want to scrutinize the isospin mass
differences of singly heavy baryons, based on the pion mean-field
approach. As mentioned previously, the light and heavy baryons are
treated on an equal footing within this approach. This means that we
can use the fixed parameters for the EM corrections to the baryon
octet, which was already done in Ref.~\cite{Yang:2010id}. These
parameters provide directly the EM corrections to the isospin mass
differences of singly heavy baryons, which come from two light
quarks inside a singly heavy baryon. Moreover, the parameters, which
are responsible for contributions from the mass difference of the up
and down quarks,  were also determined already by describing the
isospin mass differences of the baryon octet and decuplet. Therefore,
in the present work, we only need to introduce two physical
parameters, which arise from the strong hyperfine interactions between
the two light quarks inside a chiral soliton, and the EM Coulomb
interactions between the soliton and a heavy quark for both the
charmed and beauty baryons. We will select two experimental data on
the isospin mass differences of the charmed baryons to fix these two
parameters. With all theses \emph{fixed} parameters, we will show that
we describe very well the remaining isospin mass differences of the
charmed baryons. Then the isospin mass differences of the beauty
baryons are well predicted within the present framework.

The present work is organized as follows: In Section~\ref{sec:2}, we
briefly show how to compute the isospin mass differences of singly
heavy baryons from the pion mean-field approach. In addition, we
derive various mass sum rules such as the Coleman-Glashow-like 
and Guadagnini-like ones. In Section~\ref{sec:3}, we present the
numerical results for the isospin mass differences of both the
charmed and beauty baryons. We discuss in detail each contribution to
them, comparing the corresponding result with the existing
experimental data. We finally compare the present results with those
from other works.  The last Section is devoted to the summary and
conclusion of the present work.

\section{Pion mean fields and isospin symmetry breaking\label{sec:2}} 
As mentioned previously, the effects of isospin symmetry breaking are
attributed to two different sources: the EM interaction
and the difference of the up and down quark masses. Thus, we will
first consider the corrections to the masses of the singly heavy
baryons from the EM self-energies. The EM corrections to SU(3)
baryon masses were already discussed in
Ref.~\cite{Yang:2010id}. As will be shown in this work, the present
pion mean-field approach has a great virtue in dealing with the light
and heavy baryons on an equal footing also in the case of the isospin
symmetry breaking due to the EM interactions. In general, the EM
contribution to a baryon mass is expressed as ~\cite{Cottingham:1963zz} 
\begin{equation}
M_{B}^{\mathrm{EM}} \;=\; \frac{1}{2}\int d^{3}x\,d^{3}y
\langle B|T[J_{\mu}(\bm{x})J^{\mu}(\bm{y})]|B\rangle 
D_{\gamma}(\bm{x},\bm{y}) \;=\;
\langle B|\mathcal{O}^{\mathrm{EM}}|B\rangle\,,
\label{eq:corr}
\end{equation}
where $|B\rangle$ denotes the baryon state and $J^{\mu}$ the EM
current defined as 
\begin{align}
  \label{eq:EM_curr}
J^{\mu}(x)\;=\;e\bar{\psi}(x)\gamma_{\mu}\hat{\mathcal{Q}}\psi(x)\,,   
\end{align}
with the electric charge $e$ and the quark charge operator
\begin{align}
\hat{\mathcal{Q}} \;=\;
  \begin{pmatrix}
    2/3  & 0 & 0 \\ 0 & -1/3 & 0\\ 0 & 0 & -1/3 
  \end{pmatrix}
  = \frac12 \left(\lambda_3 +  \frac1{\sqrt{3}}  \lambda_8\right)\,. 
\label{eq:GN}
\end{align}
Equation~\eqref{eq:GN} is the well-known Gell-Mann-Nishijima
relation. $\lambda_3$ and $\lambda_8$ denote the Gell-mann matrices in
SU(3). The $D_{\gamma}$ represents the static photon propagator. This
is absorbed in parameters that can be fitted to the
experimental data. Since the EM current is considered as an flavor
octet operator, we write the most general form of the
$\mathcal{O}_{\mathrm{EM}}$ as a collective operator  
\begin{align}
\mathcal{O}^{\mathrm{EM}} 
 =  \alpha_{1}\sum_{i=1}^{3}
D_{Qi}^{(8)}D_{Qi}^{(8)}
+\alpha_{2}\sum_{p=4}^{7}D_{Qp}^{(8)}D_{Qp}^{(8)}
+\alpha_{3}D_{Q8}^{(8)}D_{Q8}^{(8)}\,,
\label{eq:emop}
\end{align}
where $D_{Qa}^{(8)}$ denotes the combinations of the SU(3) Wigner $D$
functions, which are defined by
\begin{align}
D_{Qa}^{(8)}(A)= \frac12 \left(D_{3a}^{(8)}(A)+\frac1{\sqrt{3}}
  D_{8a}^{(8)}(A)   \right).
\end{align}
Here, $A$ designates the rotation in flavor SU(3) space.
The parameters $\alpha_{i}$ encode specific dynamics of chiral
solitonic models and contain the photon static propagator. They were
already fixed by the empirical data on the 
EM mass differences of the baryon octet~\cite{Yang:2010id}. 
The product of two octet operators can be expanded as the SU(3)
Clebsch-Gordan (CG) series: $\mathbf{1}\oplus\mathbf{8_{s}}
\oplus\mathbf{8_{a}}\oplus\mathbf{10} \oplus
\mathbf{\overline{10}}\oplus\mathbf{27}$.   
Note, however, that because of Bose symmetry we are left only with
the singlet, the octet, and the eikosiheptaplet ($\bm{27}$), which are all
symmetric. Thus, $\mathcal{O}^{\mathrm{EM}}$ contains only the CG
series $\mathbf{1}\oplus\mathbf{8_{s}}\oplus\mathbf{27}$, written as  
\begin{align}
\mathcal{O}^{\mathrm{EM}} 
& =  
c^{(27)}\left(\sqrt{5}D_{\Sigma_{2}^{0}\Lambda_{27}}^{(27)}
+\sqrt{3}D_{\Sigma_{1}^{0}\Lambda_{27}}^{(27)}
+D_{\Lambda_{27}\Lambda_{27}}^{(27)}\right)
  +  c^{(8)}\left(\sqrt{3}D_{\Sigma^{0}\Lambda}^{(8)}
+D_{\Lambda\Lambda}^{(8)}\right)
\;+\;c^{(1)}D_{\Lambda\Lambda}^{(1)}\,.
\label{eq:emop3}
\end{align}
The explicit definitions of the Wigner $D$ functions $D_{B_1
  B_2}^{(\mathcal{R})}(A)$ can be found in Ref.~\cite{Yang:2010fm}. 
The new set of parameters $c^{(\mathcal{R})}$ can be expressed in
terms of $\alpha_i$. The superscript $(\mathcal{R})$ stands for the
corresponding irreducible representation. The last term in
Eq.~\eqref{eq:emop3} does not contribute to the mass splittings due to
isospin symmetry breaking, since the contributions with that are
cancelled out. The EM masses of singly heavy baryons can be obtained
by sandwiching the collective operator $\mathcal{O}_{\mathrm{EM}}$ in 
Eq.~\eqref{eq:emop} between the corresponding baryon states.
In fact, the parameters $c^{(\mathcal{R})}$ are slightly influenced by
changing the pion mean fields for singly heavy baryons. However,
since the EM corrections are smaller than the hadronic
contributions, i.e., the mass difference between the up and down
quarks, we can safely ignore it.  


By diagonalizing the collective Hamiltonian, we are able to derive the
collective wavefunction of a state with flavor $F=(Y,T,T_{3})$ and
spin $S=(Y'=-2/3,J,J_{3})$ in the representation $\mathcal{R}$: 
\begin{align}
\psi_{(\mathcal{R};\,F),(\overline{\mathcal{R}};\,\overline{S})}(A)
=\sqrt{\mathrm{dim}(\mathcal{R})}(-1)^{\mathcal{Q}_{B}}[D_{F\,S}^{(\mathcal{R})}(A)]^{*}\,, 
\label{eq:SolitonWF1}
\end{align}
where $\mathrm{dim}(\mathcal{R})$ denotes the dimension of the
representation $\mathcal{R}$ and $\mathcal{Q}_{B}$ a charge
corresponding to the baryon state $B$,
i.e. $\mathcal{Q}_{B}=J_{3}+Y'/2$. For a detailed formalism, we refer
to Refs.~\cite{Christov:1995vm, Blotz:1992pw}. 
To construct the complete wavefunction for a singly heavy baryon, we
need to couple the collective wavefunction to the heavy quark spinor
such that the heavy baryon state becomes a color singlet. Thus, the
wavefunction for a singly heavy baryon should be written as 
\begin{align}
\Psi_{B_{\mathrm{h}}}^{(\mathcal{R})}(A)
=\sum_{J_{3},\,J_{\mathrm{h}3}}C_{\,J,J_{3}\,J_{\mathrm{h}}\,J_{\mathrm{h}3}}^{J'\,J_{3}'}\;
  \mathbf{\chi}_{J_{\mathrm{h}3}}\;
  \psi_{(\mathcal{R};\,Y,\,T,\,T_{3})(\overline{\mathcal{R}};\,Y^{\prime},\,J,\,J_{3})}(A)\,,
\label{eq:HeavyWF}
\end{align}
where $\chi_{J_{\mathrm{h}3}}$ denote the Pauli spinors and
$C_{\,J,J_{3}\,J_{\mathrm h}\,J_{\mathrm{h}3}}^{J'\,J_{3}'}$ the CG coefficients. The
subscript $\mathrm{h}$ denotes generically a heavy 
baryon with both charmed and beauty quarks. 

Having computed the matrix elements of the collective EM operator
$\mathcal{O}^{\mathrm{EM}}$ between the collective wavefunctions for
singly heavy baryons given in Eq.~\eqref{eq:HeavyWF}, we obtain 
the EM mass corrections 
\begin{align}
M_{\Lambda_{\mathrm{h}},\,\mathrm{sol}}^{\mathrm{EM}} 
& =  \frac{1}{4}c^{(8)}\;+\;c^{(1)}\,,
\cr
M_{\Xi_\mathrm{h},\,\mathrm{sol}}^{\mathrm{EM}} 
& =  \frac{3}{4}\left(T_{3}-\frac{1}{6}\right)c^{(8)}
\;+\; c^{(1)}\,,
\label{eq:EM3bar}
\end{align}
to the baryon antitriplet, and 
\begin{align}
M_{\Sigma_\mathrm{h} ^{(\ast)},\,\mathrm{sol}}^{\mathrm{EM}}  
& =  \frac{3}{10}\left(T_{3}+\frac{1}{3}\right)c^{(8)}
\;+\; \frac{1}{9}\left(T_{3}^{2}+\frac{1}{5}T_{3}-\frac{3}{5}\right)c^{(27)}
\;+\; c^{(1)}\,,
\cr
M_{\Xi_\mathrm{h} ^{\prime(\ast)},\,\mathrm{sol}}^{\mathrm{EM}} 
& =  \frac{3}{10}\left(T_{3}-\frac{1}{6}\right)c^{(8)}
\;-\; \frac{2}{45}\left(T_{3}^{2}+2T_{3}+\frac{1}{4}\right)c^{(27)}
\;+\; c^{(1)}\,,
\cr
M_{\Omega_\mathrm{h} ^{(\ast)},\,\mathrm{sol}}^{\mathrm{EM}}  
& =  -\frac{1}{5}\left(c^{(8)}-\frac{1}{9}c^{(27)}\right)\;+\;c^{(1)}\,,
\label{eq:EMform}
\end{align}
to the baryon sextet. Note that the EM mass corrections we consider
here are only taken from the chiral soliton that consists of the light
quarks inside the singly heavy baryons. It is of great interest to see
that the eikosiheptaplet parts of
$\mathcal{O}^{\mathrm{EM}}$ do not contribute to the baryon
antitriplet. The EM corrections have the same effects on the baryon
sextet with both spin 1/2 and 3/2.
The contributions from the flavor singlet with $c^{(1)}$ are absorbed
in the classical mass of the chiral soliton. 

We now turn to the mass corrections from the isospin symmetry breaking
by the mass difference between up and down quarks. 
We call them the hadronic (H) corrections to the masses of the singly 
heavy baryons. As discussed in Ref.~\cite{Yang:2010fm}, the collective 
Hamiltonian with isospin symmetry breaking is expressed as   
\begin{align}
  H_{\mathrm{sb}}^{\mathrm{iso}}
 = 
  \left(m_{\mathrm{d}}-m_{\mathrm{u}}\right)
\left(\frac{\sqrt{3}}{2} \alpha D_{38}^{(8)}\left(R\right)
  +\beta\hat{T}_{3}
  +\frac{1}{2}\gamma\sum_{i=1}^{3}D_{3i}^{(8)}\left(R\right)\hat{J}_{i}\right)\,,
  \label{eq:Hhad}
\end{align}
where $\alpha$, $\beta$, and $\gamma$ can be fixed by using
the experimental data on the masses of the baryon octet. When it comes
to the mass spectra of the singly heavy baryons, the presence of
$N_c-1$ valence quarks instead of $N_c$ valence quarks makes the pion
mean fields undergo changes, as mentioned previously in
Introduction. While parameters $\beta$ and $\gamma$ are kept intact in
the course of changing the mean fields, $\alpha$ should be
modified. Therefore, as we did for the dynamical parameters for SU(3)
symmetry breaking in Ref.~\cite{Yang:2016qdz}, we have to replace the
$N_c$ factor by $N_c-1$, which reflects the presence of the $N_c-1$
valence quarks. Thus, we use $\overline{\alpha}$ to distinguish from
the original parameter $\alpha$. Using the results from
Ref.~\cite{Yang:2010fm}, we can immediately obtain the numerical
values of  $\overline{\alpha}$, $\beta$, and $\gamma$ as  
\begin{align}
\left(m_{\mathrm{d}}-m_{\mathrm{u}}\right)\overline{\alpha} 
& = (-2.93 \pm 0.002)\;\mathrm{MeV}\,,
\cr
\left(m_{\mathrm{d}}-m_{\mathrm{u}}\right)\beta 
& =  (-2.41\pm 0.001) \;\mathrm{MeV}\,,
\cr
\left(m_{\mathrm{d}}-m_{\mathrm{u}}\right)\gamma 
& =  (-1.74 \pm 0.006)\;\mathrm{MeV}\,.
\label{eq:abr}
\end{align}
We can redefine the parameters for each contribution to the baryon
antitriplet and sextet,
i.e. $\delta_{\boldsymbol{\overline{3}}}^{\mathrm{iso}}$ and
$\delta_{\boldsymbol{6}}^{\mathrm{iso}}$, which are explicitly written
as   
\begin{align}
  \delta_{\boldsymbol{\overline{3}}}^{\mathrm{iso}}
  & = 
  \frac{9}{16}\left(m_{d}-m_{u}\right)\left(\overline{\alpha}+\frac{16}{9}\beta\right)\,,
  \cr
  \delta_{\boldsymbol{6}}^{\mathrm{iso}}
  & = 
  \frac{9}{40}\left(m_{d}-m_{u}\right)\left(\overline{\alpha}
  +\frac{40}{9}\beta
  -\frac{4}{3}\gamma\right)\,.
  \label{eq:deltaiso}
\end{align}
Note that $\gamma$ contributes only to the baryon sextet. As displayed
in Eq.~\eqref{eq:Hhad}, the last term with $\gamma$ is related to the
spin of a singly heavy baryon. Since the light quarks in the baryon
antitriplet compose the spin $J_{\mathrm{sol}}=0$ state, i.e. a
soliton with $J_{\mathrm{sol}}=0$, the baryon antitriplet does not get
any contribution from $\gamma$. 

The baryon sextet with spin $J=1/2$ or $J=3/2$ is constructed by the
soliton with $J_{\mathrm{sol}}=1$ and a heavy quark with
$J_{\mathrm{h}}=1/2$ ($\bm{J}=
\bm{J}_{\mathrm{sol}}+\bm{J}_{\mathrm{h}}$), whereas the 
baryon antitriplet with spin $1/2$ is formed by combining the soliton
with $J_{\mathrm{sol}}=0$ and a heavy quark with
$J_{\mathrm{h}}=1/2$. This implies that the contributions of the 
strong hyperfine interactions should come into play due to the 
spin arrangement of the light quarks. In Ref.~\cite{Yang:2016qdz}, we
did not need to consider the strong hyperfine splittings between the light
quarks, since Ref.~\cite{Yang:2016qdz} investigated the mass
splittings of the singly heavy baryons only within each flavor SU(3)
representation. On the other hand, it is essential to take into
account the strong hyperfine interactions between the light quarks in
the present case, because the configuration of the light-quark spin
will definitely affect the mass of each baryon in the baryon
antitriplet and sextet, as shown in various quark
models~\cite{Chan:1985ty, Capstick:1987cw, Karliner:2019lau}.   
Thus, we introduce explicitly the strong hyperfine interaction for
the light quarks inside a heavy baryon  
\begin{align}
 H_{\mathrm{hf}} =  \delta^{\mathrm{hf}} \,\bm{S}_{1}\cdot\bm{S}_{2}\,,
  \label{eq:hyperf}
\end{align}
where $\bm{S}_1$ and $\bm{S}_2$ denote the spin operators for the
light quarks inside a soliton, which yield the soliton spin
$\bm{J}_{\mathrm{sol}}=\bm{S}_1 + \bm{S}_2$. Parameter
$\delta^{\mathrm{hf}}$ contains the masses of the up and down quarks,
and the strength of the strong hyperfine interaction. However, we will
simply fit it to the experimental data, which will be shown soon.

The EM interactions between the soliton and a heavy quark should be
involved in the isospin symmetry breaking for the masses of the singly
heavy baryons. While the magnetic interactions are suppressed by the
mass of the heavy quarks~\cite{Oka:2013xxa}, the Coulomb interactions
should be taken into account. Thus, we ignore the magnetic
interactions between the 
soliton and a heavy quark but we introduce their Coulomb
interaction of which the form is taken to be 
\begin{align}
  H_{\mathrm{sol}\mbox{-}\mathrm{h}}^{\mathrm{Coul}} 
 = 
  \alpha_{\mathrm{sol}{\mbox{-}}\mathrm{h}} \hat{Q}_{\mathrm{sol}}
  \hat{Q}_{\mathrm{h}}\,, 
  \label{eq:QsolQh}
\end{align}
where $\hat{Q}_{\mathrm{sol}}$ and $\hat{Q}_{\mathrm{h}}$ represent
the charge operators acting on the soliton and a heavy quark,
respectively. $\alpha_{\mathrm{sol}{\mbox{-}}\mathrm{h}}$ consists of
the expectation value of the inverse distance and the fine structure constant.
However, we will determine its value by adjusting it to the
experimental data.  

The parameters $\delta^{\mathrm{hf}}$ and
$\alpha_{\mathrm{sol}{\mbox{-}}\mathrm{h}}$ can be fixed by using the  
experimental data on the isospin mass differences~\cite{Tanabashi:2018oca}  :
\begin{align}
  \label{eq:fixing}
  M_{\Xi_{c}^{+}}-M_{\Xi_{c}^{0}} &= (-2.98\pm 0.22)\,\mathrm{MeV}\,, \cr
M_{\Sigma_{c}^{+}}-M_{\Sigma_{c}^{0}} &= (-0.9\pm 0.4)\,\mathrm{MeV}\,.
\end{align}
Thus, they are determined to be 
\begin{align}
\delta^{\mathrm{hf}} &=(0.40\pm 0.06)\,\mathrm{MeV}\,, \cr
\alpha_{\mathrm{sol}{\mbox{-}}\mathrm{h}} &= (2.76\pm0.28)\,\mathrm{MeV}\,.
\label{eq:para}
\end{align}
We want to emphasize that except for these two parameters we have no
free parameters at all, since all other parameters were fixed in the
light baryon sector.
\begin{table}[h]
\caption{Compilation of the formulae for each contribution to the
  isospin mass differences in the baryon antitriplet and sextet
  representations. The first column 
  denotes the irreducible representations of charmed and beauty
  baryons and the second one lists various isospin mass
  differences. In the third column, the expressions for the
  electromagnetic corrections to the corresponding isospin mass
  differences, whereas the fourth and fifth ones list the
  contributions of the up and down quark mass difference, and those of
  the strong hyperfine interactions, respectively. The last column
  lists the corrections from the Coulomb interactions between the
  soliton and a heavy quark.} 
\global\long\def\arraystretch{1.5}%

\begin{tabular}{c|ccccc}
\hline 
    {$\boldsymbol{\mathcal{R}_{J}}$}
    & $B_\mathrm{h}$
    & {$\Delta M_{\mathrm{sol}}^{\mathrm{EM}}$}
    & {$\Delta M_{\mathrm{sb}}^{\mathrm{iso}}$}
    & {$\Delta M_{\mathrm{hf}}$}
    & {$\Delta M_{\mathrm{sol}{\mbox{-}}\mathrm{h}}^{\mathrm{Coul}}$}
    \tabularnewline
\hline 
\multirow{2}{*}{{$\boldsymbol{\overline{3}}_{1/2}$}}
& {$\Xi_{c}^{+}-\Xi_{c}^{0}$}
& \multirow{2}{*}{{${\displaystyle \frac{3}{4}c^{\left(8\right)}}$}}
& \multirow{2}{*}{{$\delta_{\boldsymbol{\overline{3}}}^{\mathrm{iso}}$}}
& \multirow{2}{*}{{$-3\delta^{\mathrm{hf}}$}}
& {$2\alpha_{\mathrm{sol}{\mbox{-}}\mathrm{h}}$}
\tabularnewline
& {$\Xi_{b}^{0}-\Xi_{b}^{-}$}
&
&
&
& {$-\alpha_{\mathrm{sol}{\mbox{-}}\mathrm{h}}\,/\,3$}
\tabularnewline
\hline 
\multirow{10}{*}{{$\boldsymbol{6}_{1/2\left(3/2\right)}$}}
& {$\Sigma_{c}^{\ast++}-\Sigma_{c}^{\ast+}$}
& \multirow{2}{*}{{${\displaystyle
    \frac{3}{10}\left(c^{\left(8\right)}+
  \frac{4}{9}c^{\left(27\right)}\right)}$}}
& \multirow{2}{*}{{$\delta_{\boldsymbol{6}}^{\mathrm{iso}}$}}
& \multirow{2}{*}{{$\delta^{\mathrm{hf}}$}}
& {$2\alpha_{\mathrm{sol}{\mbox{-}}\mathrm{h}}$}
\tabularnewline
& {$\Sigma_{b}^{\ast+}-\Sigma_{b}^{\ast0}$}
&
&
&
& {$-\alpha_{\mathrm{sol}{\mbox{-}}\mathrm{h}}\,/\,3$}
\tabularnewline
\cline{2-6} \cline{3-6} \cline{4-6} \cline{5-6} \cline{6-6} 
& {
    $
    \begin{array}{c}
       \Sigma_{c}^{\ast+}-\Sigma_{c}^{\ast0}
    \end{array}
    $}
& \multirow{2}{*}{{${\displaystyle
         \frac{3}{10}\left(c^{\left(8\right)}
         -\frac{8}{27}c^{\left(27\right)}\right)}$}}
& \multirow{2}{*}{{$\delta_{\boldsymbol{6}}^{\mathrm{iso}}$}}
& \multirow{2}{*}{{$\delta^{\mathrm{hf}}$}}
& {$2\alpha_{\mathrm{sol}{\mbox{-}}\mathrm{h}}$}
\tabularnewline
& {$\Sigma_{b}^{\ast0}-\Sigma_{b}^{\ast-}$}
&
&
&
& {$-\alpha_{\mathrm{sol}{\mbox{-}}\mathrm{h}}\,/\,3$}
\tabularnewline
\cline{2-6} \cline{3-6} \cline{4-6} \cline{5-6} \cline{6-6} 
& {$\Xi_{c}^{\prime\ast+}-\Xi_{c}^{\prime\ast0}$}
& \multirow{2}{*}{{${\displaystyle
 \frac{3}{10}\left(c^{\left(8\right)} -\frac{8}{27}c^{\left(27\right)}\right)}$}}
& \multirow{2}{*}{{$\delta_{\boldsymbol{6}}^{\mathrm{iso}}$}}
& \multirow{2}{*}{{$\delta^{\mathrm{hf}}$}}
& {$2\alpha_{\mathrm{sol}{\mbox{-}}\mathrm{h}}$}
\tabularnewline
& {$\Xi_{b}^{\prime\ast0}-\Xi_{b}^{\prime\ast-}$}
&
&
&
& {$-\alpha_{\mathrm{sol}{\mbox{-}}\mathrm{h}}\,/\,3$}
\tabularnewline
\cline{2-6} \cline{3-6} \cline{4-6} \cline{5-6} \cline{6-6} 
& {$\Sigma_{c}^{\ast++}-\Sigma_{c}^{\ast0}$}
  & \multirow{2}{*}{{${\displaystyle \frac{3}{5}\left(c^{\left(8\right)}
    +\frac{2}{27}c^{\left(27\right)}\right)}$}}
& \multirow{2}{*}{{$2\delta_{\boldsymbol{6}}^{\mathrm{iso}}$}}
& \multirow{2}{*}{{$2\delta^{\mathrm{hf}}$}}
& {$2\alpha_{\mathrm{sol}{\mbox{-}}\mathrm{h}}$}
\tabularnewline
& {$\Sigma_{b}^{\ast+}-\Sigma_{b}^{\ast-}$}
&  &  &  & {$-\alpha_{\mathrm{sol}{\mbox{-}}\mathrm{h}}\,/\,3$}
\tabularnewline
\cline{2-6} \cline{3-6} \cline{4-6} \cline{5-6} \cline{6-6} 
  & {$\Sigma_{c}^{\ast++}+\Sigma_{c}^{\ast0}
    -2\Sigma_{c}^{\ast+}$}
& \multirow{2}{*}{{${\displaystyle \frac{2}{9}c^{\left(27\right)}}$}}
& \multirow{2}{*}{{$0$}}
& \multirow{2}{*}{{$0$}}
& \multirow{2}{*}{{$0$}}
\tabularnewline
  & {$\Sigma_{b}^{\ast+}+\Sigma_{b}^{\ast-}
    -2\Sigma_{b}^{\ast0}$}
&  &  &  &
\tabularnewline
\hline 
\end{tabular}{\small\par}

\label{tab:1}
\end{table}
In Table~\ref{tab:1}, we compile the formulae for each contribution to
the isospin mass differences in the baryon antitriplet and sextet 
representations.  

It is of great interest to examine the mass sum rules or the relations
between the isospin mass differences of the singly heavy baryons. The
present work provide various relations among isospin multiplets.  
When $\Delta T_{3}=1$, we find the following relations
\begin{align}
M_{\Sigma_{c}^{++}}-M_{\Sigma_{c}^{+}} & =
 M_{\Sigma_{c}^{\ast++}}-M_{\Sigma_{c}^{\ast+}}\,, 
\label{eq:sum0}\\
M_{\Sigma_{b}^{+}}-M_{\Sigma_{b}^{0}} 
& = M_{\Sigma_{b}^{\ast+}}-M_{\Sigma_{b}^{\ast0}}\,.
\label{eq:sum1}
\end{align}
In Eq.~\eqref{eq:sum0} $M_{\Sigma_{c}^{++}}-M_{\Sigma_{c}^{+}}$ and
$M_{\Sigma_{c}^{\ast++}} -M_{\Sigma_{c}^{\ast+}}$ belong to the baryon
sextet with spin 1/2 and 3/2, respectively. In Eq.~\eqref{eq:sum1} we
gives the same relation for the beauty sextet. We can obtain another
relation for the isospin mass differences of the charmed baryons in
the sextet as follows: 
\begin{align}
\underbrace{\begin{array}{c}
      M_{\Sigma_{c}^{+}} -
              M_{\Sigma_{c}^{0}}\end{array}}_{\text{\normalsize$(-0.9\pm0.4)\,\mathrm{MeV}$}} 
 = \begin{array}{c} M_{\Sigma_{c}^{\ast+}} - M_{\Sigma_{c}^{\ast0}}\end{array}
\;=\;
\underbrace{ M_{\Xi_{c}^{\prime+}} -
  M_{\Xi_{c}^{\prime0}}}_{\text{\normalsize$(-0.8\pm0.6)\,\mathrm{MeV}$}} 
\;=\;
\underbrace{ M_{\Xi_{c}^{\ast+}} -
  M_{\Xi_{c}^{\ast0}}}_{\text{\normalsize$(-0.80\pm0.26)\,\mathrm{MeV}$}}\,. 
\label{eq:sum2}
\end{align}
Having inserted the experimental values into each term in
Eq.~\eqref{eq:sum2}, we find that the relation is in good agreement
with the data. We also observe that for $\Delta T_3=1$ the
$\Sigma_{b}$ and $\Xi_{b}$ multiplets in the sextet satisfy the
following relation: 
\begin{align}
  M_{\Sigma_{b}^{0}} - M_{\Sigma_{b}^{-}}
   =  \;\; M_{\Sigma_{b}^{\ast0}} - M_{\Sigma_{b}^{\ast-}}
  \;\;=\;\;
  M_{\Xi_{b}^{\prime0}} - M_{\Xi_{b}^{\prime-}}
  \;\;=\;\;
  M_{\Xi_{b}^{\ast0}} - M_{\Xi_{b}^{\ast-}}\,,
  \label{eq:sum3}
\end{align}
which are simply the same as that in Eq.~\eqref{eq:sum2} except for
the heavy-quark flavor. The relations given in Eqs.~\eqref{eq:sum2}
and~\eqref{eq:sum3} remind us the Coleman-Glashow mass
formula~\cite{Coleman:1961jn} that relates the isospin mass
differences of the baryon octet. As for the $\Delta T_3=2$, we obtain
the relations for the baryon sextet with both spin 1/2 and 3/2: 
\begin{align}
  \underbrace{ M_{\Sigma_{c}^{++}} -
  M_{\Sigma_{c}^{0}}}_{\mathrm{\text{\normalsize$(0.220\pm0.013)\,\mathrm{MeV}$}}} 
  & \;= \;
  \underbrace{ M_{\Sigma_{c}^{\ast++}} -
    M_{\Sigma_{c}^{\ast0}}}_{\mathrm{\text{\normalsize$(0.01\pm0.15)\,\mathrm{MeV}$}}}\,, 
  \label{eq:sum4}\\
  \underbrace{ M_{\Sigma_{b}^{+}} -
  M_{\Sigma_{b}^{-}}}_{\mathrm{\text{\normalsize$(-5.06\pm0.18)\,\mathrm{MeV}$}}} 
  & \;= \; 
  \underbrace{ M_{\Sigma_{b}^{\ast+}} -
    M_{\Sigma_{b}^{\ast-}}}_{\mathrm{\text{\normalsize-$(4.37\pm0.33)\,\mathrm{MeV}$}}}\,.  
  \label{eq:sum5}
\end{align}
We put the experimental values to check the relations. As shown in the
underbraces, it seems that they deviate slightly from the experimental
data.

There is also a neutral mass sum rule~\cite{Franklin:1975yu}, in which
each sum of the heavy baryon masses have the neutral charge: 
\begin{align}
  &(M_{\Sigma_{c}^{++}} + M_{\Sigma_{c}^{0}} - 2M_{\Sigma_{c}^{+}})
   = 
  (M_{\Sigma_{c}^{\ast++}} + M_{\Sigma_{c}^{\ast0}} -
    2M_{\Sigma_{c}^{\ast+}}) 
\cr
  = 
 &\;(  M_{\Sigma_{b}^{+}} + M_{\Sigma_{b}^{-}} - 2M_{\Sigma_{b}^{0}})
  = 
(  M_{\Sigma_{b}^{\ast+}} + M_{\Sigma_{b}^{\ast-}}-2M_{\Sigma_{b}^{\ast0}})
\cr
 = &\;\frac{1}{2} (M_{\Sigma^{++}} + M_{\Sigma^{0}} - 2M_{\Sigma^{+}})
 \approx \frac{3}{2}(M_{\Sigma^{\ast ++}} + M_{\Sigma^{\ast 0}} - 2M_{\Sigma^{\ast +}})
\,.
  \label{eq:sum6}
\end{align}
The neutral mass sum rule in Eq.~\eqref{eq:sum6} is usually given for
the charmed baryons. However, we can extend this sum rule by including
the beauty baryons as shown in Eq.~\eqref{eq:sum6}. The
neutral mass sum rule has a unique feature. All the hadronic
contributions are cancelled each other, so that only the EM
corrections remain~\cite{Gasser:1982ap}. Thus, Eq.~\eqref{eq:sum6} can
be considered as the EM mass relation. As will be explicitly shown
later, the neutral mass sum rule in Eq.~\eqref{eq:sum6} is in
remarkable agreement with the experimental data. 
It implies that the present treatment for the EM contributions,
which was determined already in the light baryon sector, describes
universally well both the EM isospin mass differences of light and
heavy baryons.  

The Guadagnini mass formula~\cite{Guadagnini:1983uv} relates the
mass difference in the baryon decuplet to that in the
octet. Similarly, we find that the isospin mass differences in the
baryon antitriplet can be related to those in the baryon sextet as follows: 
\begin{align}
  \underbrace{\left(M_{\Xi_{c}^{+}} - M_{\Xi_{c}^{0}}\right)
    \;-\; \left(M_{\Xi_{b}^{0}} -
  M_{\Xi_{b}^{-}}\right)}_{\mathrm{\text{\normalsize$(2.92\pm0.64)\,\mathrm{MeV}$}}} 
  & =  \frac{1}{2}\left[\left(M_{\Sigma_{c}^{(\ast)++}} - M_{\Sigma_{c}^{\ast0}}\right)
  \;-\;           \left(M_{\Sigma_{b}^{(\ast)+}} - M_{\Sigma_{b}^{\ast-}}\right)\right]
  \cr
  & =  \underbrace{\frac{1}{2}\left[\left(M_{\Sigma_{c}^{++}} -
    M_{\Sigma_{c}^{0}}\right) 
    \;-\;    \left(M_{\Sigma_{b}^{+}} -
    M_{\Sigma_{b}^{-}}\right)\right]}_{\mathrm{\text{\normalsize$(2.64\pm0.09)\,\mathrm{MeV}$}}} 
  \cr
  & =  \underbrace{\frac{1}{2}\left[\left(M_{\Sigma_{c}^{\ast++}} -
    M_{\Sigma_{c}^{\ast0}}\right) 
    \;-\;   \left(M_{\Sigma_{b}^{\ast+}} -
    M_{\Sigma_{b}^{\ast-}}\right)\right]}_{\mathrm{\text{\normalsize$(2.19\pm0.18)
    \,\mathrm{MeV}$}}}    
  \cr
  & =  \underbrace{\frac{1}{2}\left[\left(M_{\Sigma_{c}^{++}} -
    M_{\Sigma_{c}^{0}}\right) 
    \;-\;   \left(M_{\Sigma_{b}^{\ast+}} -
    M_{\Sigma_{b}^{\ast-}}\right)\right]}_{\mathrm{\text{\normalsize$(2.29\pm0.17)\,
    \mathrm{MeV}$}}}   
  \cr
  & =  \underbrace{\frac{1}{2}\left[\left(M_{\Sigma_{c}^{\ast++}} -
    M_{\Sigma_{c}^{\ast0}}\right) 
    \;-\;   \left(M_{\Sigma_{b}^{+}} -
    M_{\Sigma_{b}^{-}}\right)\right]}_{\mathrm{\text{\normalsize$(2.53\pm0.12)
    \,\mathrm{MeV}$}}}   
  \cr
  & =  \left(M_{\Xi_{c}^{\prime\ast+}} - M_{\Xi_{c}^{\prime\ast0}}\right)
  \;-\;\left(M_{\Xi_{b}^{\prime\ast0}} - M_{\Xi_{b}^{\prime\ast-}}\right)\,,
  \label{eq:sum7}
\end{align}
which are well satisfied. We can also obtain two more mass relations
among the baryon sextet:
\begin{align}
  \left(M_{\Sigma_{c}^{\ast++}} - M_{\Sigma_{c}^{\ast+}}\right)
  \;+\;\left(M_{\Xi_{b}^{\prime\ast0}} - M_{\Xi_{b}^{\prime\ast-}}\right)
  & =  \frac{1}{2}\left[\left(M_{\Sigma_{c}^{\ast++}} - M_{\Sigma_{c}^{\ast0}}\right)
  \;+\;\left(M_{\Sigma_{b}^{\ast+}} - M_{\Sigma_{b}^{\ast-}}\right)\right]\,,
  \label{eq:sum9}\\
 \left( M_{\Sigma_{b}^{\ast+}} - M_{\Sigma_{b}^{\ast-}}\right)
  & =  \left(M_{\Sigma_{b}^{\ast+}} - M_{\Sigma_{b}^{\ast0}}\right)
  \;+\;\left(M_{\Xi_{b}^{\prime\ast0}} - M_{\Xi_{b}^{\prime\ast-}}\right).
  \label{eq:sum8}
\end{align}

\section{Results and discussion \label{sec:3}}
We are now in a position to discuss the results from the present
work. 
\begin{table}[htp]
\caption{Isospin mass differences of the charmed baryon antitriplet
  and sextet in units of MeV. The first column shows the SU(3)
  representations of the singly heavy baryons. The second one
  indicates the isospin mass difference in a given representation. The
  third one lists the results of the electromagnetic corrections of
  the light quarks to the mass of a singly heavy baryon, $\Delta
  M_{\mathrm{sol}}^{EM}$.  The fourth one gives the corrections from
  the mass difference of the up and down quarks, $\Delta
  M_{\mathrm{sb}}^{\mathrm{iso}}$. The fifth one lists the corrections
  from the strong hyperfine interactions between the light quarks,
  $\Delta M_{\mathrm{hf}}$. The sixth one presents the Coulomb
  interactions between the soliton and a heavy quark, $\Delta
  M_{\mathrm{sol}{\mbox{-}}\mathrm{h}}^{\mathrm{Coul}}$. The seventh
  one lists the total results of the isospin mass differences of the
  charmed baryon antitriplet and sextet. The eighth
  one lists the corresponding experimental data on the isospin mass
  differences taken from  the Particle Data Group
  (PDG)~\cite{Tanabashi:2018oca}. In the last column, we list the
  derived values of the isospin mass differences, using the
  experimental data on the masses of the corresponding singly heavy
  baryons.}      
\global\long\def\arraystretch{1.5}%
\begin{raggedright}
{\small{}}%
\begin{tabular}{ccccccccc}
\hline 
$\boldsymbol{\mathcal{R}_{J}}$ 
& $ B_{c}$ 
& {$\Delta M_{\mathrm{sol}}^{\mathrm{EM}}$}
& {$\Delta M_{\mathrm{sb}}^{\mathrm{iso}}$}
& {$\Delta M_{\mathrm{hf}}$}
& {$\Delta M_{\mathrm{sol}{\mbox{-}}\mathrm{h}}^{\mathrm{Coul}}$}
& $\Delta M^{\mathrm{total}}$ 
& PDG \cite{Tanabashi:2018oca}
& PDG${}^{\dagger}$
\tabularnewline
\hline 
$\boldsymbol{\overline{3}}_{1/2}$ 
& $\Xi_{c}^{+}-\Xi_{c}^{0}$ 
& $-0.11\pm0.17$ 
& $-3.51$ 
& $-1.20\pm0.18$ 
& $1.84\pm0.19$ 
& input & $-2.98\pm0.22$ 
& $-$ 
\tabularnewline
\hline 
 & $\Sigma_{c}^{++}-\Sigma_{c}^{+}$ 
& $1.10\pm0.33$ 
& $-2.33$ 
& $0.40\pm0.06$ 
& $1.84\pm0.19$ 
& $1.02\pm0.38$ 
& $-$ 
& $ 1.07 \pm 0.42 $
\tabularnewline
$\boldsymbol{6}_{1/2}$ 
& $\begin{array}{c}
\Sigma_{c}^{+}-\Sigma_{c}^{0}\end{array}$ 
& $-0.81\pm0.22$ 
& $-2.33$ 
& $0.40\pm0.06$ 
& $1.84\pm0.19$ 
& input 
& $-0.9\pm0.4$
& $-$
\tabularnewline
 & $\Xi_{c}^{\prime+}-\Xi_{c}^{\prime0}$ 
& $-0.81\pm0.22$ 
& $-2.33$ 
& $0.40\pm0.06$ 
& $1.84\pm0.19$ 
& $-0.90\pm0.30$ 
& $-0.8\pm0.6$
& $-$
\tabularnewline
 & $\Sigma_{c}^{++}-\Sigma_{c}^{0}$ 
& $0.29\pm0.17$ 
& $-4.66$ & $0.80\pm0.12$ 
& $3.68\pm0.37$ 
& $0.12\pm0.43$ 
& $0.220\pm0.013$ 
& $-$
\tabularnewline
 & $\Sigma_{c}^{++}+\Sigma_{c}^{0}-2\Sigma_{c}^{+}$ 
& $1.92\pm0.53$ 
& $0$ & $0$ & $0$ 
& $1.92\pm0.53$ 
& $-$
& $1.92 \pm 0.82$
\tabularnewline
\hline 
 & $\Sigma_{c}^{\ast++}-\Sigma_{c}^{\ast+}$ 
& $1.10\pm0.33$ 
& $-2.33$ 
& $0.40\pm0.06$ 
& $1.84\pm0.19$ 
& $1.02\pm0.38$ 
& $-$
& $0.91 \pm 2.31$
\tabularnewline
$\boldsymbol{6}_{3/2}$ 
& $\Sigma_{c}^{\ast+}-\Sigma_{c}^{\ast0}$ 
& $-0.81\pm0.22$ 
& $-2.33$ 
& $0.40\pm0.06$ 
& $1.84\pm0.19$ 
& $-0.90\pm0.30$ 
& $-$
& $-0.98 \pm 2.31$
\tabularnewline
 & $\begin{array}{c}
\Xi_{c}^{\ast+}-\Xi_{c}^{\ast0}\end{array}$ 
& $-0.81\pm0.22$ 
& $-2.33$ 
& $0.40\pm0.06$ 
& $1.84\pm0.19$ 
& $-0.90\pm0.30$ 
& $-0.80\pm0.26$
& $-$
\tabularnewline
 & $\Sigma_{c}^{\ast++}-\Sigma_{c}^{\ast0}$ 
& $0.29\pm0.17$ 
& $-4.66$ 
& $0.80\pm0.12$ 
& $3.68\pm0.37$ 
& $0.12\pm0.43$ 
& $0.01\pm0.15$
& $-$
\tabularnewline
 & $\Sigma_{c}^{\ast++}+\Sigma_{c}^{\ast0}-2\Sigma_{c}^{\ast+}$ 
& $1.92\pm0.53$ 
& $0$ 
& $0$ 
& $0$ 
& $1.92\pm0.53$ 
& $-$
& $1.89 \pm 4.64$
\tabularnewline
\hline 
\end{tabular}
\end{raggedright}
\label{tab:2}
\end{table} 
In Table~\ref{tab:2}, we list the results of each contribution to the
isospin mass differences of the charmed baryon antitriplet and sextet.
We want to emphasize again that except for the contributions from the
strong hyperfine interactions, $\Delta M_{\mathrm{hf}}$, and the Coulomb
interactions between the soliton and a heavy quark, $\Delta
M_{\mathrm{sol}{\mbox{-}}\mathrm{h}}^{\mathrm{Coul}}$, all the other
terms were determined by using the parameters fixed already in the
light baryon sectors.  Comparing the values of
$\Delta_{\mathrm{sol}}^{\mathrm{EM}}$ with the experimental data that
are listed in the eighth and last columns, we interestingly observe
that the EM corrections from the light quarks describe almost 
quantitatively the experimental data apart from the baryon
antitriplet. It indicates that the hadronic contributions from the
mass difference of the up and down quarks, $\Delta
M_{\mathrm{sb}}^{\mathrm{iso}}$, are almost cancelled out by both the
strong hyperfine interactions, $\Delta_{\mathrm{hf}}$, and the Coulomb
interactions between the soliton and a heavy quark,
$\Delta
M_{\mathrm{sol}{\mbox{-}}\mathrm{h}}^{\mathrm{Coul}}$. However, the
small amounts that are left after the cancellations are rather
important to describe the experimental data quantitatively. On the
other hand, the effects of the hadronic part that includes
$m_{\mathrm{d}}-m_{\mathrm{u}}$ are the most dominant ones.
The total results are generally in good agreement with the
experimental data. As explained previously, the neutral mass
differences, $M_{\Sigma_c^{++}}+M_{\Sigma_c^0} - 2 M_{\Sigma_c^+}$ and 
$M_{\Sigma_c^{*++}}+M_{\Sigma_c^{*0}} - 2 M_{\Sigma_c^{*+}}$, only
contain the EM contributions. We emphasize that all the parameters
have been fixed already by reproducing the EM mass differences for the
baryon octet.  Moreover, the present results are in remarkably good
agreement with the experimental data, as shown in
Table~\ref{tab:2}. This implies that the present pion mean-field
approach indeed explains consistently both the SU(3) light baryons
and the singly heavy baryons. 

\begin{table}[htp] 
\caption{Isospin mass differences of the beauty baryon antitriplet
  and sextet in units of MeV. The notations are the same as in
  Table~\ref{tab:2}.} 
{\small{}}%
\begin{raggedright}
\begin{tabular}{ccccccccc}
\hline 
$\boldsymbol{\mathcal{R}_{J}}$ 
& $ B_{b}$ 
& {$\Delta M_{\mathrm{sol}}^{\mathrm{EM}}$}
& {$\Delta M_{\mathrm{sb}}^{\mathrm{iso}}$}
& {$\Delta M_{\mathrm{hf}}$}
& {$\Delta M_{\mathrm{sol}{\mbox{-}}\mathrm{h}}^{\mathrm{Coul}}$}
& $\Delta M^{\mathrm{total}}$ 
& PDG \cite{Tanabashi:2018oca}
& PDG${}^{\dagger}$
\tabularnewline
\hline 
$\boldsymbol{\overline{3}}_{1/2}$ 
& $\Xi_{b}^{0}-\Xi_{b}^{-}$ 
& $-0.11\pm0.17$ 
& $-3.51$ 
& $-1.20\pm0.18$ 
& $-0.92\pm0.09$ 
& $-5.74\pm0.27$ 
& $-5.9\pm0.6$
& $-$
\tabularnewline
\hline 
 & $\Sigma_{b}^{+}-\Sigma_{b}^{0}$ 
& $1.10\pm0.33$ 
& $-2.33$ 
& $0.40\pm0.06$ 
& $-0.92\pm0.09$ 
& $-1.74\pm0.34$ 
& $-$
& $-$
\tabularnewline
$\boldsymbol{6}_{1/2}$ 
& $\Sigma_{b}^{0}-\Sigma_{b}^{-}$ 
& $-0.81\pm0.22$ 
& $-2.33$ 
& $0.40\pm0.06$ 
& $-0.92\pm0.09$ 
& $-3.66\pm0.25$ 
& $-$
& $-$
\tabularnewline
 & $\Xi_{b}^{\prime0}-\Xi_{b}^{\prime-}$ 
& $-0.81\pm0.22$ 
& $-2.33$ 
& $0.40\pm0.06$ 
& $-0.92\pm0.09$ 
& $-3.66\pm0.25$ 
& $-$
& $-$
\tabularnewline
 & $\Sigma_{b}^{+}-\Sigma_{b}^{-}$ 
& $0.29\pm0.17$ 
& $-4.66$ 
& $0.80\pm0.12$ 
& $-1.84\pm0.19$ 
& $-5.40\pm0.28$ 
& $-5.06\pm0.18$
& $-$
\tabularnewline
 & $\Sigma_{b}^{+}+\Sigma_{b}^{-}-2\Sigma_{b}^{0}$ 
& $1.92\pm0.53$ 
& $0$ 
& $0$ 
& $0$ 
& $1.92\pm0.53$ 
& $-$
\tabularnewline
\hline 
 & $\Sigma_{b}^{\ast+}-\Sigma_{b}^{\ast0}$ 
& $1.10\pm0.33$ 
& $-2.33$ 
& $0.40\pm0.06$ 
& $-0.92\pm0.09$ 
& $-1.74\pm0.34$ 
& $-$
& $-$
\tabularnewline
$\boldsymbol{6}_{3/2}$ 
& $\Sigma_{b}^{\ast0}-\Sigma_{b}^{\ast-}$ 
& $-0.81\pm0.22$ 
& $-2.33$ 
& $0.40\pm0.06$ 
& $-0.92\pm0.09$ 
& $-3.66\pm0.25$ 
& $-$
& $-$
\tabularnewline
 & $\Xi_{b}^{\ast0}-\Xi_{b}^{\ast-}$ 
& $-0.81\pm0.22$ 
& $-2.33$ 
& $0.40\pm0.06$ 
& $-0.92\pm0.09$ 
& $-3.66\pm0.25$ 
& $-$
& $-3.03\pm0.91$
\tabularnewline
 & $\Sigma_{b}^{\ast+}-\Sigma_{b}^{\ast-}$ 
& $0.29\pm0.17$ 
& $-4.66$ 
& $0.80\pm0.12$ 
& $-1.84\pm0.19$ 
& $-5.40\pm0.28$ 
& $-4.37\pm0.33$
&$-$
\tabularnewline
 & $\Sigma_{b}^{\ast+}+\Sigma_{b}^{\ast-}-2\Sigma_{b}^{\ast0}$ 
& $1.92\pm0.53$ 
& $0$ 
& $0$ 
& $0$ 
& $1.92\pm0.53$ 
& $-$
& $-$
\tabularnewline
\hline 
\end{tabular}{\small\par}
\par\end{raggedright}
\label{tab:3}
\end{table}
Table~\ref{tab:3} lists the results of each contribution to the
isospin mass differences of the beauty baryon antitriplet and
sextet. Since we have already fixed the parameters for the strong
hyperfine and Coulomb interactions, the results listed in
Table~\ref{tab:3} are the predictions of the present work. Note that
there are only three experimental data. In addition we can extract one
more data by using the experimental values of the $\Xi_b^{*0}$ and
$\Xi_b^{*-}$ masses. The present results describe the experimental
data very well. For example, the value of the isospin mass difference
of the beauty baryon antitriplet is in very good agreement with the 
corresponding data by approximately $3\,\%$. This implies that 
heavy-quark flavor symmetry is well satisfied. Note that in the
present work, we do not have any contribution from heavy quark
mass corrections such as chromoelectric or chromomagnetic interactions
that are proportional to $1/m_{\mathrm{h}}$. 

\begin{figure}[htp]
  \raggedright{}
  \includegraphics[scale=1.6]{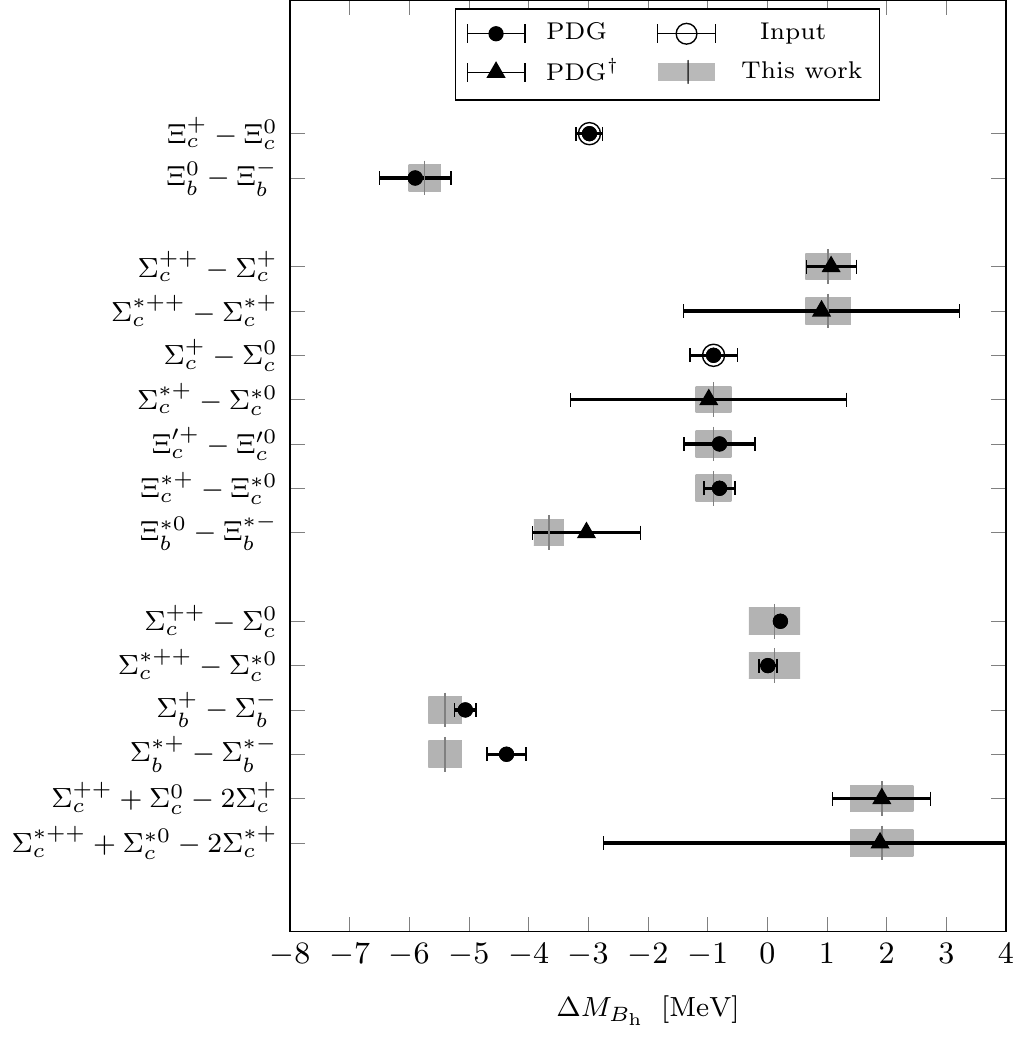}
\caption{Comparison of the present results with the corresponding
  experimental data. The $x$-axis denotes the values of the isospin
  mass differences between singly heavy baryons in units of MeV
  whereas the $y$-axis designates the corresponding mass difference of
the isospin multiplet. The filled circles stand for the experimental
data taken from the PDG~\cite{Tanabashi:2018oca}, the filled triangles
represent the data obtained by using the experimental values of the
masses of the corresponding heavy baryons~\cite{Tanabashi:2018oca},
and the open circles designate the data taken to be as input. The
shaded rectangles represent the present results.}  
  \label{fig:1}
\end{figure}
To compare the present results with the experimental data more
visibly, we illustrate in Fig.~\ref{fig:1} the results for the 
isospin mass difference of both the charmed and beauty baryons, which
we have already discussed in Tables~\ref{tab:2} and~\ref{tab:3}.

\begin{sidewaystable}[H]
\caption{Comparison of the present results for the isospin mass
  differences of the charmed baryon antitriplet and sextet with those 
  from a quark model (QM), a relativized quark model (RQM),
  the semirelativistic quark model (SRQM), a constituent quark model
  (CQM), a potential model (PM), the chiral bag model ($\chi$BM), and
  the MIT bag model in units of MeV. The numerical result of the mass 
  difference for   $\Xi_{c}^{\prime+}-\Xi_{c}^{\prime0}$ from chiral perturbation
  theory \cite{Guo:2008ns} is given as $(-0.2\pm0.6)\,\mathrm{MeV}$.}   
\renewcommand{\arraystretch}{1.2}
{}%
\hspace{-4em}
\begin{tabular}{ccccccccccccccccc}
\hline 
$\boldsymbol{\mathcal{R}_{J}}$
& $ B_{c}$
& PDG \cite{Tanabashi:2018oca}
& PDG${}^{\dagger}$
& This work
& QM\cite{Chan:1985ty}
& QM\cite{Lichtenberg:1977mv}
& QM\cite{Karliner:2008sv}
& RQM\cite{Capstick:1987cw}
& SRQM\cite{Itoh:1988xs}
& CQM\cite{Verma:1988gg}
& CQM\cite{Cutkosky:1993cc}
& PM\cite{Varga:1998wp}
& PM\cite{SilvestreBrac:2003kd}
& $\chi$BM\cite{Hwang:1986ee}
& MIT\cite{Tiwari:1985ru}
\tabularnewline
\hline 
$\boldsymbol{\overline{3}}_{1/2}$
& $\Xi_{c}^{+}-\Xi_{c}^{0}$
& $-2.98\pm0.22$
& $-3.00\pm0.43$
& $\mathrm{input}$
& $-3.20$
& $0.6$
& $-3.1\pm0.5$
& $ - $
& $ - $
& $-0.9$
& $ - $
& $-2.83$
& $-3.42$
& $-2.1$
& $-1.72$\tabularnewline
\hline 
& $\Sigma_{c}^{++}-\Sigma_{c}^{+}$
& $ - $
& $ 1.07\pm0.42$
& $1.02\pm0.38$
& $1.05$
& $2.6$
& $ - $
& $ - $
& $ - $
& $1.2$
& $ - $
& $1.56$
& $ - $
& $3.5$
& $0.82$
\tabularnewline
$\boldsymbol{6}_{1/2}$
& $\begin{array}{c}
  \Sigma_{c}^{+}-\Sigma_{c}^{0}\end{array}$
& $-0.9\pm0.4$
& $ -0.85\pm0.42$
& $\mathrm{input}$
& $-0.73$
& $0.8$
& $ - $
& $-0.2$
& $-0.35$
& $ - $
& $ - $
& $-0.36$
& $-0.33$
& $ - $
& $-0.83$
\tabularnewline
& $\Xi_{c}^{\prime+}-\Xi_{c}^{\prime0}$
& $-0.8\pm0.6$
& $ -0.80\pm0.71$
& $-0.90\pm0.30$
& $-1.00$
& $-1.1$
& $-2.3\pm4.24$
& $ - $
& $-0.34$
& $-1.2$
& $-0.52\pm0.33$
& $-0.30$
& $-0.20$
& $-1.0$
& $-1.48$
\tabularnewline
& $\Sigma_{c}^{++}-\Sigma_{c}^{0}$
& $0.220\pm0.013$
& $0.22\pm0.20$
& $0.12\pm0.43$
& $-0.05$
& $ - $
& $ - $
& $1.4$
& $1.08$
& $0.5$
&$ - $
& $1.20$
& $0.37$
& $3.0$
& $ - $
\tabularnewline
& $\Sigma_{c}^{++}+\Sigma_{c}^{0}-2\Sigma_{c}^{+}$
& $ - $
& $1.92\pm0.82 $
& $1.92\pm0.53$
& $ - $
& $ - $
& $ - $
& $ - $
& $ - $
& $ - $
& $ - $
& $ - $
& $ - $
& $ - $
& $ - $
\tabularnewline
\hline 
& $\Sigma_{c}^{\ast++}-\Sigma_{c}^{\ast+}$
& $ - $
& $ 0.91\pm2.31$
& $1.02\pm0.38$
& $0.81$
& $1.7$
& $ - $
& $ - $
& $ - $
& $0.9$
& $ - $
& $ - $
& $ - $
& $3.3$
& $0.63$
\tabularnewline
$\boldsymbol{6}_{3/2}$
& $\Sigma_{c}^{\ast+}-\Sigma_{c}^{\ast0}$
& $ - $
& $ -0.98\pm2.31$
& $-0.90\pm0.30$
& $-0.97$
& $-0.1$
& $ - $
& $-0.8$
& $-1.21$
& $-0.9$
& $ - $
& $ - $
& $ - $
& $ - $
& $-0.95$
\tabularnewline
 & $\begin{array}{c}
  \Xi_{c}^{\ast+}-\Xi_{c}^{\ast0}\end{array}$
& $-0.80\pm0.26$
& $ -0.81\pm0.33$
& $-0.90\pm0.30$
& $-1.24$
& $-0.3$
& $0.5\pm1.84$
& $ - $
& $-1.08$
& $-1.2$
& $-0.44\pm0.37$
& $-0.43$
& $-0.25$
& $-1.3$
& $-0.86$
\tabularnewline
& $\Sigma_{c}^{\ast++}-\Sigma_{c}^{\ast0}$
& $0.01\pm0.15$
& $ -0.07\pm0.29$
& $0.12\pm0.43$
& $0.02$
& $ - $
& $ - $
& $-0.1$
& $-0.29$
& $ - $
& $1.00\pm0.52$
& $ - $
& $0.19$
& $2.7$
& $ - $
\tabularnewline
& $\Sigma_{c}^{\ast++}+\Sigma_{c}^{\ast0}-2\Sigma_{c}^{\ast+}$
& $ - $
& $ 1.89\pm4.61$
& $1.92\pm0.53$
& $ - $
& $ - $
& $ - $
& $ - $
& $ - $
& $ - $
& $ - $
& $ - $
& $ - $
& $ - $
& $ - $
\tabularnewline
\hline 
\end{tabular}
\label{Tab:resultCharm}
\vspace{3em}
\caption{Comparison of the present results for the isospin mass
  differences of the beauty baryon antitriplet and sextet with those
  from the quark model (QM), heavy-quark effective theory (HQET),
  chiral perturbation theory ($\chi$PT), a relativized quark model
  (RQM), the semirelativistic quark model (SRQM), a potential model
  (PM), and the chiral bag model ($\chi$BM) in units of MeV.} 
\renewcommand{\arraystretch}{1.5}

{}%
\begin{tabular}{cccccccccccc}
\hline 
$\boldsymbol{\mathcal{R}_{J}}$
& $ B_{b}$
& PDG \cite{Tanabashi:2018oca}
& PDG${}^{\dagger}$
& This work
& QM\cite{Chan:1985ty}
& HQET\cite{Hwang:2008dj}
& $\chi$PT\cite{Guo:2008ns}
& RQM\cite{Capstick:1987cw}
& SRQM\cite{Itoh:1988xs}
& PM\cite{Varga:1998wp}
& $\chi$BM\cite{Hwang:1986ee}
\tabularnewline
\hline 
$\boldsymbol{\overline{3}}_{1/2}$
& $\Xi_{b}^{0}-\Xi_{b}^{-}$
& $-5.9\pm0.6$
& $-5.10\pm1.03 $
& $-5.74\pm0.27$
& $-6.16$
& $-6.9\pm1.1$
& $ - $
& $ - $
& $ - $
& $-5.39$
& $ - $
\tabularnewline
\hline 
& $\Sigma_{b}^{+}-\Sigma_{b}^{0}$
& $ - $
& $ -$
& $-1.74\pm0.34$
& $-2.17$
& $ - $
& $ - $
& $ - $
& $ - $
& $ - $
& $-1.5$
\tabularnewline
$\boldsymbol{6}_{1/2}$
& $\Sigma_{b}^{0}-\Sigma_{b}^{-}$
& $ - $
& $ - $
& $-3.66\pm0.25$
& $-3.95$
& $-4.7\pm1.0$
& $-4.9\pm1.9$
& $-3.7$
& $-2.74$
& $-2.51$
& $ - $
\tabularnewline
& $\Xi_{b}^{\prime0}-\Xi_{b}^{\prime-}$
& $ - $
& $- $
& $-3.66\pm0.25$
& $-4.21$
& $ - $
& $-4.0\pm1.9$
& $ - $
& $-2.74$
& $ - $
& $ - $
\tabularnewline
& $\Sigma_{b}^{+}-\Sigma_{b}^{-}$
& $-5.06\pm0.18$
& $ -5.08\pm0.37$
& $-5.40\pm0.28$
& $-6.07$
& $ - $
& $ - $
& $-5.6$
& $-3.70$
& $-3.57$
& $-7.1$
\tabularnewline
& $\Sigma_{b}^{+}+\Sigma_{b}^{-}-2\Sigma_{b}^{0}$
& $ - $
& $ -$
& $1.92\pm0.53$
& $ - $
& $ - $
& $ - $
& $ - $
& $ - $
& $ - $
& $ - $
\tabularnewline
\hline 
& $\Sigma_{b}^{\ast+}-\Sigma_{b}^{\ast0}$
& $ - $
& $ - $
& $-1.74\pm0.34$
& $-2.02$
& $ - $
& $ - $
& $ - $
& $ - $
& $ - $
& $-1.2$\tabularnewline
$\boldsymbol{6}_{3/2}$
& $\Sigma_{b}^{\ast0}-\Sigma_{b}^{\ast-}$
& $ - $
& $ -$
& $-3.66\pm0.25$
& $-3.80$
& $ - $
& $ - $
& $-3.6$
& $-3.21$
& $ - $
& $ - $
\tabularnewline
& $\Xi_{b}^{\ast0}-\Xi_{b}^{\ast-}$
& $ - $
& $-3.03\pm0.91 $
& $-3.66\pm0.25$
& $-4.01$
& $ - $
& $ - $
& $ - $
& $-3.08$
& $ - $
& $ - $\tabularnewline
& $\Sigma_{b}^{\ast+}-\Sigma_{b}^{\ast-}$
& $-4.37\pm0.33$
& $ -4.42\pm0.40$
& $-5.40\pm0.28$
& $-5.85$
& $ - $
& $ - $
& $-5.4$
& $-4.30$
& $ - $
& $-6.5$
\tabularnewline
& $\Sigma_{b}^{\ast+}+\Sigma_{b}^{\ast-}-2\Sigma_{b}^{\ast0}$
& $ - $
& $ -$
& $1.92\pm0.53$
& $ - $
& $ - $
& $ - $
& $ - $
& $ - $
& $ - $
& $ - $
\tabularnewline
\hline 
\end{tabular}
\label{Tab:resultbottom}
\end{sidewaystable}

\section{Summary and outlook \label{sec:4}}
In the present work, we investigated the isospin mass differences of
the singly heavy baryons within the framework of a pion mean-field
approach.  Since there are two different sources that break the
isospin symmetry, i.e., the electromagnetic interactions and the mass
difference of the up and down quarks, we need to treat them
within the same framework. We started with computing the
electromagnetic contributions to the masses of the baryon antitriplet
and sextet. Because of bose symmetry of photons, we have only
symmetric contributions such as the SU(3) singlet, octet, and the
eikosiheptaplet ($\bm{27}$). This means that we have three independent
parameters, of which the singlet terms can be absorbed in the
classical soliton mass. The remaining two parameters can be fixed by
using the empirical data on the electromagnetic mass differences of
the baryon octet. The hadronic contributions from the mass difference
of the up and down quarks can be taken into account perturbatively.
These contributions were also fixed by reproducing the masses of the
baryon octet. Thus, we have no free parameters related to the
light-quark sector. In addition, we introduced the strong hyperfine
interactions of the light quarks inside a soliton, which play an
essential role in describing the different spin configuration of the baryon
antitriplet from the baryon sextet. In doing so, we have one free
parameter to be fixed by using the experimental data on the
isospin mass differences of the charmed baryons.    
While a heavy quark inside a singly heavy baryon
is regarded as a mere static color source, we still need to consider
the electromagnetic Coulomb interactions between the soliton and a
heavy quark. It brings an additional free parameter, which will be
also fixed by using the experimental data as done with that of the
strong hyperfine interactions.  Once we fixed these two free
parameters, we were able to produce all possible isospin mass
differences of both the charmed and beauty baryons.
We derived sum rules among the isospin mass differences
of the singly heavy baryons, which are similar to the Coleman-Glashow
sum rules. The pion mean-field approach has one great virtue, which
relates the isospin mass differences of the baryon antitriplet to
those of the baryon sextet. These mass relations resemble the
Guadagnini mass formula. These sum rules are in good agreement with
the experimental data. In addition, we also obtained the mass
relations among the members of the baryon sextet.

With the two parameters fixed by using two of the experimental data, we 
proceeded to computing the isospin mass differences of the charmed
baryons. We observed that the electromagnetic contributions from the
light quarks described already the experimental data on the isospin
mass differences of the baryon sextet almost quantitatively. Having
included all the contributions, we showed that the present results are
in very good agreement with the experimental data. Interestingly, the
neutral mass relations for the charmed baryons $\Sigma_c$
($\Sigma_c^*$), which are written as $M_{\Sigma_{c}^{++}} +
M_{\Sigma_{c}^{0}} - 2M_{\Sigma_{c}^{+}}$ ($M_{\Sigma_{c}^{*++}} +
M_{\Sigma_{c}^{*0}} - 2M_{\Sigma_{c}^{*+}}$), contain only the
electromagnetic interactions of  the light quarks. This indicates that
the neutral mass relations provide a stringent test on the present
framework. Indeed, the results of the neutral sum rules are in
remarkable agreement with the experimental data. This implies that the
present pion mean-field approach explains consistently both the
isospin mass differences of the light baryons and singly heavy
baryons. Finally, we compared the present results with those from
various different works.  

\section*{Acknowledgments}
The present work was supported by Basic Science Research Program through
the National Research Foundation of Korea funded by the Ministry of
Education, Science and Technology (NRF-2019R1A2C1010443 (Gh.-S. Y.),
2018R1A2B2001752, and 2018R1A5A1025563 (H.-Ch.K.)).

\begin{appendix}
  \section{Relations of the electromagnetic mass differences}
For completeness, we present various relations of the EM mass
differences as    
\begin{align}
  \Delta M\left[\Sigma_{c}^{\ast++}
    -\Sigma_{c}^{\ast0}\right]_{\mathrm{sol}}^{\mathrm{EM}} 
& =\;\;
  \Delta M\left[\Sigma_{c}^{\ast++}
    -\Sigma_{c}^{\ast+}\right]_{\mathrm{sol}}^{\mathrm{EM}}
  \;+\;\Delta M\left[\Sigma_{c}^{\ast+}
    -\Sigma_{c}^{\ast0}\right]_{\mathrm{sol}}^{ \mathrm{EM}}
\cr
\;\;=\;\;
\Delta M\left[\Sigma_{b}^{\ast+}
  -\Sigma_{b}^{\ast-}\right]_{\mathrm{sol}}^{ \mathrm{EM}} 
& =\;\;
\Delta M\left[\Sigma_{b}^{\ast+}
  -\Sigma_{b}^{\ast0}\right]_{\mathrm{sol}}^{ \mathrm{EM}}
\;+\;\Delta M\left[\Sigma_{b}^{\ast0}
  -\Sigma_{b}^{\ast-}\right]_{\mathrm{sol}}^{\mathrm{EM}}\,,\cr
  \Delta M\left[\Sigma_{c}^{\ast+}
    -\Sigma_{c}^{\ast0}\right]_{\mathrm{sol}}^{ \mathrm{EM}} 
& =\;\;
  \Delta M\left[\Xi_{c}^{\prime\ast+}
    -\Xi_{c}^{\prime\ast0}\right]_{\mathrm{sol}}^{\mathrm{ EM}}\,,
\cr
\;\;\;\;\;\;
\Delta M\left[\Sigma_{b}^{\ast0}
  -\Sigma_{b}^{\ast-}\right]_{\mathrm{sol}}^{ \mathrm{EM}} 
& =\;\;
\Delta M\left[\Xi_{b}^{\prime\ast0}
  -\Xi_{b}^{\prime\ast-}\right]_{\mathrm{sol}}^{ \mathrm{EM}}\,,\cr
  \Delta M\left[\Xi_{c}^{+}-\Xi_{c}^{0}\right]_{\mathrm{sol}}^{ \mathrm{EM}}
  & =\;\;\Delta M\left[\Sigma_{c}^{\ast++}
    -\Sigma_{c}^{\ast0}\right]_{\mathrm{sol}}^{ \mathrm{EM}}
  \;+\;\frac{1}{2}\Delta M\left[\Sigma_{c}^{\ast+}
    -\Sigma_{c}^{\ast0}\right]_{\mathrm{sol}}^{ \mathrm{EM}}
  \cr
  \;\;=\;\;\Delta M\left[\Xi_{b}^{0}
    -\Xi_{b}^{-}\right]_{\mathrm{sol}}^{ \mathrm{EM}}
  & =\;\;\Delta M\left[\Sigma_{b}^{\ast+}
    -\Sigma_{b}^{\ast-}\right]_{\mathrm{sol}}^{ \mathrm{EM}}
  \;+\;\frac{1}{2}\Delta M\left[\Sigma_{b}^{\ast0}
    -\Sigma_{b}^{\ast-}\right]_{\mathrm{sol}}^{ \mathrm{EM}}\,.
\label{eq:EMrel3}
\end{align}
Note that these relations show the heavy-quark flavor symmetry in the
limit of $m_Q\to\infty$. 
\end{appendix}

\end{document}